# Learning to Collide: Recommendation System Model Compression with Learned Hash Functions

Benjamin Ghaemmaghami, Mustafa Ozdal, Rakesh Komuravelli, Dmitriy Korchev, Dheevatsa Mudigere, Krishnakumar Nair, Maxim Naumov. 2022

## Abstract

A key characteristic of deep recommendation models is the immense memory requirements of their embedding tables. These embedding tables can often reach hundreds of gigabytes which increases hardware requirements and training cost. A common technique to reduce model size is to hash all of the categorical variable identifiers (ids) into a smaller space. This hashing reduces the number of unique representations that must be stored in the embedding table; thus decreasing its size. However, this approach introduces "collisions" between semantically dissimilar ids that degrade model quality. We introduce an alternative approach, Learned Hash Functions, which instead learns a new mapping function that encourages collisions between semantically similar ids. We derive this learned mapping from historical data and embedding access patterns. We experiment with this technique on a production model and find that a mapping informed by the combination of access frequency and a learned low dimension embedding is the most effective. We demonstrate a small improvement relative to the hashing trick and other collision related compression techniques. This is ongoing work that explores the impact of categorical id collisions on recommendation model quality and how those collisions may be controlled to improve model performance.

## Introduction

Recommendation systems are a fundamental data center workload underlying personalization, ranking, and content selection. One fundamental component of these workloads are large embedding tables that contain trained multi-dimensional representations of categorical features. Each embedding table typically contains tens of millions of vectors, where each vector has 10s or 100s of dimensions. Furthermore, a single model can contain several hundreds of such tables, each corresponding to a different categorical feature. As a result; sizes of the state-of-the-art models can reach multiple terabytes [Naumov et al., 2020]. However, hardware has not caught up with model size needs: the available high bandwidth memory (HBM) on a state-of-the-art GPU is typically in the range of 32-80GBs [NVIDIA, 2021]. Efficient training of a large model requires distributing the embedding tables across many GPUs, which incurs increased communication overhead and training time. As recommendation system model complexity has been doubling every two years [Naumov et al., 2020], techniques to reduce memory footprint will become more and more essential. Recommendation systems represent a significant part of modern data center workloads [Gupta et al., 2020] and model size improvements provide the opportunity to run higher quality models on the same infrastructure. In order to continue scaling deep learning recommender systems, we need new techniques that will allow us to reduce model size in both training and inference.

A core goal of this work is to develop a technique that reduces model size in both training and inference: an "end-to-end" size reduction. The most helpful model compression techniques are the ones that allow more complicated models to run on existing infrastructure, thus end-to-end reductions in model size are very important. An example of an end-to-end reduction is the use of quantization to reduce numerical bit-width across training and inference.

In deep learning based recommendation systems, an embedding table is a mapping from categorical context variables to a real valued multi-dimensional representation [Naumov et al., 2019]. The height of an embedding table, for a given feature, is equal to the number of unique categorical values the feature can assume. The width of an embedding table is the dimension of the trained representation. Each feature can represent information like topics, videos liked, or other contextual data. Embedding tables have two knobs to control model size: 1) the dimension of the vectors (width) and 2) the number of unique context values mapped by the table (height). We can easily reduce the dimension of the vectors by adjusting a hyperparameter, but reducing the height of the table is more complicated. The most basic technique to achieve height reduction is to restrict the space of context variables. Let's use an example of a model that incorporates the user's favorite color and favorite movie genre as context. Reducing the height of the favorite color table could mean restricting the values within to red, green, or blue. Fundamentally, this is not an issue as we have only lost resolution of the color. However, in the case of favorite movie genres, how do we determine which genre can be removed?

One common solution to reduce table height is to randomly hash all of the unique context values into a shorter table [Weinberger et al., 2010]. With hashing, we are still able to represent all topics, but some topics will share embeddings. More formally, this "hashing trick" changes the mapping implemented by the embedding table from one-to-one to onto. A pair of context values that map to the same embedding are said to "collide" These collisions are problematic as the trained embedding will now be influenced by both values. What if the hash function maps two context values that appeal to opposite groups to the same embedding? The hash function has no information about how these pages relate to each other: the job of training is to determine that relation! In practice, these collisions can reduce model quality and make randomly hashing to a smaller table size ineffective when quality loss is not acceptable (examples of this quality degradation are shown in Figure 1).

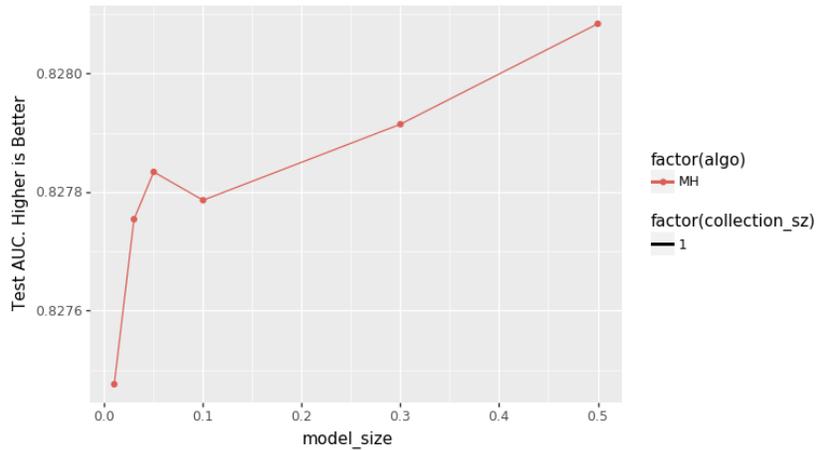

Figure 1: Model size vs quality for the hashing trick on 1x. Note that the magnitude of the Y axis is small: ~0.001

In this work we propose a technique to learn hash functions based on semantic information about the context values. Specifically, the contributions of this work are:
- A new approach to learn hash functions based on historical data
- A high performance library for embedding table clustering algorithms

# Related Work

The hashing trick [Weinberger et al., 2010], is a simple yet effective technique to reduce the height of embedding tables. Instead of having a one-to-one mapping between context values and embedding vectors, the hashing trick assigns each context value an id and hashes it into a smaller space. A frequently used hashing function is the modulo operator. This approach results in collisions between context values, but the simplicity of the approach has made it a frequently used technique.

Complementary partition based embedding tables [Shi et al., 2020] improves upon the hashing trick by constructing unique embeddings from sets of shared embeddings. Instead of performing a single embedding lookup, this technique uses lookups to multiple embedding tables and reduces these intermediate embeddings to a single vector. By using hash functions with certain properties, the complementary partitions approach can ensure that there is a unique embedding for every context value. However, while the final embeddings may be unique, the intermediate embeddings that the final embedding is constructed from still collide with other embeddings. Further, these hash functions use no information about the relation between intermediate embeddings and have the same collision problems.

Following the approach of composing embeddings from other smaller components, TT-Rec [Yin, et al., 2021] constructs embeddings using the tensor train decomposition [Oseledets, 2011]. TT-Rec also implements an uncompressed cache for frequently accessed embeddings as the authors find that this helps preserve model quality relative to the baseline.

Mixed dimension embeddings [Ginart et al., 2019] proposes modifying the embedding dimension for each embedding. The embedding dimension is chosen based upon the access frequency of the context value and more parameters are allocated to more frequently accessed values. Each embedding is unique, and parameter allocation is weighted based on frequency for each context value. This technique does not modify the number of embedding vectors in the embedding table.

Deep Hash Embeddings [Kang et al., 2021] describes an alternative recommendation model architecture that replaces embedding tables with MLPs and a novel categorical id encoding strategy. Each categorical value is encoded into a continuous vector and then this vector is passed into an MLP which produces the final embedding.

Frequency Based Double Hashing [Zhang et al., 2020] describes a technique which combines the ideas of compositional embeddings with access frequency information. This approach applies compositional embeddings to "low frequency" indices, while "high frequency" indices are isolated such that they do not collide with any other indices.

# Background: Deep Recommendation Systems

Embedding tables are used across a variety of domains, but we will focus on their use in recommendation systems. While this work is focused on production models, Deep Learning Recommendation Model (DLRM) [Naumov et al., 2019] is representative of the recommendation systems used in production at Facebook [Park et al., 2018]. DLRM is designed for click-through-rate and ranking tasks and predicts the probability of a user clicking a proposed piece of content.

Many recommendation models are built around collaborative filtering (CF): a set of techniques that recommend items to a user based on the preferences of other similar users [Aggarwal, 2016]. One version of CF is matrix factorization where the ratings matrix, the matrix of items to recommendation value, is factored into a user matrix and an item matrix. This factorization produces a low rank approximation of the full ratings matrix. Recommendations are produced by "interacting", often computing the dot product of a pair of rows from the user and item matrices. Deep recommendation models build on an extension of this factorization technique called Factorization Machines. Factorization Machines generalize the previous matrix factorization to an unbounded set of context variables [Rendle, 2010]. In this context, each context variable is assigned a separate matrix where each row corresponds to a different value. Like matrix factorization, the Factorization Machine interacts with the values from each matrix to produce a final output. To extend this to the deep context, an embedding is trained for each context variable, embeddings are interacted, and interaction outputs are passed through a set of fully-connected layers before producing a prediction. Embeddings and dense features can be processed before interaction, if so desired. There are a variety of recommendation models ([Cheng et al., 2016][Guo et al., 2017][Lian et al., 2018][Naumov et al., 2019]) that use a Deep Factorization Machine.

# The problem with collisions

Many embedding compression techniques hash categorical ids into a smaller space: resulting in collisions between those ids. Empirically, this has been shown to negatively impact model quality. Avoiding or minimizing collisions is a common idea to improve model quality, and the behavior of the hashing trick motivates this thinking. For example, if we apply the hashing trick with a simple modulo function to reduce model size by 50% we see a drop in model quality.. Figure 1 shows this relationship plotted. However, the number of collisions is not the only factor that impacts model quality. The authors of Frequency Based Double Hashing ([Zheng et al., 2020]) also noticed this behavior and used frequency information to reduce collisions for frequently accessed values.

These results indicate that quality is dictated by both the number of collisions between ids and the significance of each collision. A more general question follows - "how can we estimate the significance of each collision?" Our hypothesis is that categorical values that are more similar will have less significant collisions and thus less impact on model quality. Thus, we should be able to create a smaller model at similar quality by generating a hash function that hashes similar categorical values to the same index. The following sections show how we find similarity information and derive hash functions from it.

# Estimating categorical value similarity from low dimension latent features

A driving hypothesis for this work is that similar context values should have a shared embedding vector, but how do we decide what is similar? Determining similarity is a hard problem to solve as we know very little information about the relationship between context values before the model is trained. For example, we may have two locations: San Francisco, California, and Austin, Texas, but we do not know the impact that these locations will have on the recommendation result. They could produce similar recommendations, but we have no way to tell before the model is trained. The goal is to find some process that allows us to cluster context values such that similar values are in the same cluster.

The fact that these clusters may overlap is essential. For example, San Francisco may be similar to Austin in some contexts, but it is also similar to New York City in others. Elaborating on the example, San Francisco and Austin can be clustered by pleasant weather, and San Francisco and New York City can be clustered by population density. Clustering in only one dimension would allow us to represent only one aspect of similarity. On the other hand, multiple sets of overlapping clusters can represent the multiple relationships that exist between context values.

There are a few ways that we could use observable information about the data to cluster context values directly. For example, we could determine which context values commonly co-occur in queries or directly cluster based on some domain specific knowledge (i.e physical distance from each other for our location example). Data mining the dataset in this manner may work, but it requires significant prepossessing that exists outside of a normal model training workflow and has extreme combinatorial complexity. Additionally, domain specific approaches do not scale well to the large number of context variables used in modern recommendation models.

An alternative to data mining queries is to train the full size model once and then cluster the embeddings in the trained tables by value. After training, the trained embeddings exist in a learned latent feature space. Consider any model that contains embeddings; training this model will implicitly produce a latent feature for each of the columns in the trained embedding table. These latent feature values are not directly interpretable, but provide a similarity metric between vectors in the embedding table. Clustering in this feature space is an intuitive way to divide context values. Unfortunately, training with the full size model has high memory requirements and is unhelpful in a model compression pipeline.

In order to meet our goal of an "end-to-end" model size reduction, we propose to train a smaller version of the model to use for clustering. By drastically reducing the embedding dimension of the full size model to $D = 2$ or $D = 4$, we can create a memory-size friendly model that will still produce latent features that we can use for clustering. We do not decrease the height of the embedding tables in the low dimension model; thus the clustering discovered is applicable for all context values. Any produced clustering can be saved for future use and be reused while it remains relevant. Additionally, while training the smaller version of the model we can also collect access frequency information for each of the context values and incorporate it into our proposed clustering algorithm.

## Incorporating Category Access Frequency Information

One problem associated with using learned weights from a recommendation system is that in practice the access pattern to the weights is frequently power law distributed. An impact of this is that some weights may have only been updated a handful of times while others may have been updated millions. Even with adaptive learning rates, this leads to a quality disparity between embeddings. The infrequently updated weights will be less useful for clustering as they are much closer to randomly distributed. To compensate for this effect we propose two solutions discussed below. Each solution requires us to have a count of all accesses to every categorical value across training. We evaluate the impact of access frequency information with an ablation study in the experiments section.

### Prioritizing clustering by frequency

Our first technique to address frequency issues is to prioritize the merging of low frequency clusters first. We compute a score for each pair of neighboring clusters by first normalizing all access counts to a 1-2 range and then multiplying neighbors together. The intuition here is that the lowest frequency clusters are randomly distributed so they may be safely merged with their neighbors whether high frequency or low frequency, but the merging of low frequency clusters with other low frequency clusters is preferred.

### Weighting merged cluster values by frequency

The second way that we address access frequency while merging clusters is to weight the values of merged clusters by access frequency when producing the value for a new cluster. The idea here is to have more random points (ones with lower access frequency) influence the position of merged custers less.

# An algorithm for clustering categorical ids based on similarity, cluster size, and access frequency

Now that we have explained how we produce similarity information, we can now discuss our clustering algorithm. We propose an agglomerative clustering algorithm that operates on 1D spaces and merges a fixed number of clusters per step. Agglomerative clustering algorithms are "bottom-up;" that is, initially all values are in independent clusters and each step of the algorithm merges two clusters together. Because our problem sizes are very large (100s of tables each with millions of vectors) we implement a more generalized agglomerative algorithm that merges multiple clusters at each step. For the following description, we will refer to a potential cluster created by merging clusters together as a "candidate cluster."

In the following description, we use the notation "XY -> Z" to denote the merging of clusters X and Y into the candidate cluster Z. We also use the notation <A,B,C> to denote a sorted vector of clusters A, B and C. Given that this is a 1D clustering algorithm, the only valid merges are between neighbors or sequences of neighbors; i.e: AB->X, BC->X, or ABC->X.

The core goal of the clustering process is to cluster items that are close together in the embedding space. Each iteration of the algorithm will select a pair of clusters to merge. For example: if we have the clusters <A,B,C,D,E>, the algorithm may first elect to merge AB -> A' yielding A'CDE. We generalize this to K clusters removed on each step, so if K=2, the algorithm could elect to merge: AB -> A' and CD -> C' on the same step yielding A'C'E. The algorithm only considers pairs of neighboring clusters as sorted by value. We choose this restriction so that ranking generation is $O(N)$ instead of $O(K*N^2)$ for all possible subsequences of K clusters. We may still merge a sequence of candidate clusters into a single cluster, but the result is not explicitly considered when scoring candidates. For example if AB->A' and BC->B' are the highest scoring candidates, this sequence will be merged into ABC -> A' yielding A'DE; however, the candidate ABC -> A' is never scored.

On each iteration, the algorithm attempts to select K clusters to merge. This is done by progressively filtering the set of possible candidates. At the start of each step, all neighboring pairs are considered valid candidates, and the algorithm applies three filters to restrict the candidates: 1) frequency information, 2) cluster value, and 3) cluster size. Frequency information and cluster value both select the top K scoring values on their respective filter steps. Cluster size either filters out all candidates that have a size greater than the target, or, if all candidates are greater than the target, the cluster size filter returns the K candidates with the smallest created cluster sizes.

The filtering process is as follows:
1. 2*K candidates with the smallest normalized frequency scores
2. 1.5*K candidates with the smallest distance between clusters
3. Up to K candidates that are either smaller than the size target OR the K smallest candidates available

This process can be reordered to prioritize specific behaviors or stages can be removed entirely as shown in our experiments section.

Equation 1: Pairwise distance between clusters
$$distance_{ij} = value(c_j) - value(c_i)$$

Equation 2: Pairwise size of the combination of two clusters
$$|c_i'| = |c_i| + |c_j|$$

Equation 3: Normalized count value for each cluster
$$norm\_counts(c\_i) = (counts(c\_i) / ||C||\_2) + 1$$

$C$: *the set of all clusters*
$c_i$: *cluster i*
$value(c_i)$: *the value of cluster $c_i$*
$|c_i|$: *the size of cluster $c_i$*

Invariant 1: The set of clusters is sorted by value
$$value(c_j) \geq value(c_i)$$

The pseudocode for the algorithm is as follows. Note that the actual implementation of this algorithm is optimized for GPU and thus differs from this pseudocode.

```
1   sorted_clusters = sort(C)
2   while |sorted_clusters| > target:
3     num_to_merge = min(2*K, 2*(target - |sorted_clusters|))
4
5     norm_c_score_heap = min_heap()
6     for i in |sorted_clusters|-2:
7       j = i+1
8       ci = sorted_clusters[i]
9       cj = sorted_clusters[j]
10      norm_c_heap.insert((norm_counts(ci) * norm_counts(cj), ci, cj))
11
12    to_merge_by_distance = min_heap()
13    for i in num_to_merge:
14      _, ci, cj = frequency_score_heap.heap_pop()
15      to_merge_by_distance.insert(distance(ci,cj), ci, cj)
16
17    canidates_less_than_max = []
18    to_merge_by_size = min_heap()
19    for i in (num_to_merge * 0.75):
20      _, ci, cj = to_merge_by_distance.heap_pop()
21      canidate_size = |ci| + |cj|
22      if canidate_size < size_max:
23        canidates_less_than_max.append((ci, cj))
24      else:
25        to_merge_by_size.insert(canidate_size, ci, cj)
26
27    good_single_lookahead_choices = |canidates_less_than_max| > 0
28    valid_canidates_single_lookahead = []
29    if good_single_lookahead_choices:
30      valid_canidates_single_lookahead = canidates_less_than_max
31    else:
32      for i in (num_to_merge * 0.5):
33        _, ci, cj = to_merge_by_size.heap_pop()
34        valid_canidates_single_lookahead.append(
35          (ci, cj)
36        )
37
38    if good_single_lookahead_choices:
```

```
39        valid_canidates = split_all_combined_canidates_greater_than_max(
40          valid_canidates_single_lookahead, size_max
41        )
42    else:
43      valid_canidates = valid_canidates_single_lookahead
44
45    sorted_clusters = merge_clusters(sorted_clusters, valid_canidates)
```

## Rules for merging clusters

When merging two or more clusters together, there are a few cases we must consider: 1) how is the value of the clusters merged?, 2) how is the size of the clusters merged, and 3) how are the access counts of the clusters merged. When merging values we take the arithmetic mean of the merged clusters. When merging counts and sizes we take the sum. The count for the merged cluster is defined as the sum of the counts of the clusters that are merged. Thus a low frequency cluster merged with a high frequency cluster will continue to be considered a high frequency cluster after the merge.

Equation 4: Candidate cluster value update

$$value(c_i') = \frac{\sum_{j=i}^{N} value(c_j)}{N}$$

Equation 5: Candidate cluster size update

$$size(c_i') = \sum_{j=i}^{N} size(c_j)$$

Equation 6: Candidate cluster counts update

$$count(c_i') = \sum_{j=i}^{N} count(c_j)$$

## Removing chains of clusters that exceed maximum size

When calculating the size of candidate clusters, the pairwise approach has a drawback: chains of candidate clusters can exist that when merged will exceed the maximum size. For example, if given initial clusters ABCD each of size one and a size limit of two, the merges AB->A', BC->B' and CD->C' are all valid. However, if two (or more) of these merges are selected on the same step, the resulting cluster will be greater than the size limit.

**Problem**:

*Given* clusters={ABCD}, where |A|,|B|,|C|,|D| = 1
AB->A', |A'| = 2
*and*
BC->B', |B'| = 2
*Unify merges in the same step*
ABC->A', |A'| = 3

A simple solution to this problem is to unify the merges, check if any unifications exceed the size limit, and then break the merges into non-contiguous sets of merges. The current implementation does not take the relative sizes of the component merges into account when selecting which to split, but this is a logical extension.

**Solution:**
*Given* clusters {ABCDE}, max size =2, |A|,|B|,|C|,|D|, |E| = 1
*Choose*
AB->A' *and* BC->B' *and* CD->C' *and* DE->D'
*Unify merges*
ABCDE->A', |A'| = 5
*Breaking contiguous groups into pairs left to right*
AB->A' *and* CD->C' *and* E->E

# Implementation implications

In order to implement our approach we developed two libraries: 1) a high performance embedding table lookup with support for our custom hash functions and 2) a collection of algorithms to generate the hash functions. In this section we will discuss important details of each implementation.

## Flexible Embedding Table Lookups:

In order to support our idea and comparison to related work, we developed a high performance gpu embedding table implementation built on top of the existing FBGEMM library. While the default embedding table implementation in common libraries such as Pytorch or Tensorflow is a kernel with a single embedding table lookup, repeating such kernel calls tens or hundreds of times incurs significant overhead. Instead, we build on top of the combined embedding table kernel in FBGEMM which merges all embedding table lookups into a set of fused kernels.

To implement our learned hash functions we pair each embedding table with a lookup table that maps each input index to a row in the embedding table. Given that we may use multiple hash functions, a single *logical* embedding table may have multiple *internal* tables to service each of the hash functions. Embeddings retrieved from these internal tables are merged into a single embedding by either summation or concatenation. Gradients during backpropagation are also expanded as required to correctly update the internal tables. Input indices are expected to be in a predefined range, if this technique is required to be used in an environment where input indices frequently change those indices should be hashed into the respective range before the embedding lookup. The hash functions stored in the lookup tables have a memory usage of O(H*N) elements. The bit width of values in the lookup tables is set by the largest embedding table height.

# Experiments

We performed a variety of offline experiments with a production model and dataset. We ran all experiments on systems with server grade Intel CPUs and Nvidia GPUs (32GB V100s). The following section describes our configurations in more detail.

**Model**

Our experiments were conducted with a representative production model with an original size of O(100)GB. The model architecture is similar to DLRM as described in [Naumov et al., 2019].

**Dataset**

We experimented with a production ads dataset with *eight* consecutive days of data. We referred to those days of data as day one through day eight. From these eight days we selected three splits: 1) the training dataset for the teacher, 2) the training dataset for the clustered model, and 3) the evaluation dataset. Each data split was as follows:

- The teacher training dataset was 750M random samples from day one to day seven.
- The clustered model training dataset was the first 1.5B samples from day five.
- The evaluation dataset was the first 150M samples from day eight.

The first day of data is from a Thursday which gives the following day-of-week alignment for the dataset: Teacher Training: Thursday -> Wednesday. Clustered Training: Monday, Eval: Thursday.

**Embedding Table Model Parameters**

The following table contains information about the model size and embedding table makeup.

Table 1. Model Embedding Table Properties

| Metric | O(X) |
|---|---|
| Number of Tables | O(100) |
| Largest Table | O(10M) |

| Median Table Height | O(1M) |
|---|---|
| 25th Percentile Height | O(10K) |
| 75th Percentile Height | O(10M) |
| Table Dim Range | O(1) - O(100) |

**Hyperparameters**

We used the same hyperparameters for all of our experiments. We used sparse rowwise adagrad as the optimizer for the embedding tables and LAMB as the optimizer for the MLPs, a learning rate for the embedding tables of 0.02, a learning rate for the MLPs of 0.002, and a batch size of 32768. We used synchronous gradient updates during training. We do not reduce the height of tables with size less than 100K vectors as doing so has minimal impact on model size, but has significant impact on model quality as shown in Figure 2 below.

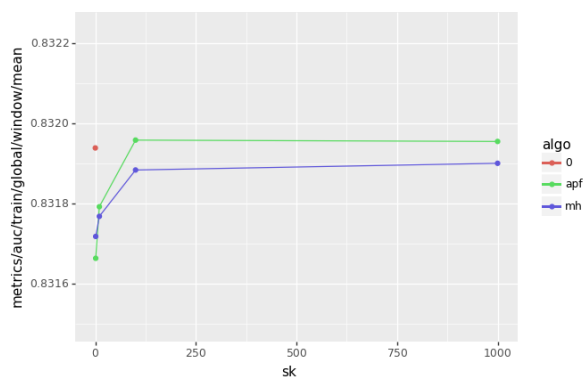

Figure 2: Impact of skipping small tables on model quality. APF is agglomerative clustering, MH is modulo hashing, and 0 is the baseline. SK is the cutoff value for applying the hashing techniques in thousands of rows. For example, SK 1000 means that only tables with 1M or more rows had hashing techniques applied. We see that changing the cutoff from tables with less than 100K to all tables significantly decreases model quality for both our approach and the hashing trick.

**Metric**

We repeat each experiment at least two times and report results in terms of mean AUC.

To evaluate our technique we answer the following research questions:
1. Does our technique learn to avoid unconstructive collisions?
2. To what degree does the addition of access frequency information improve quality?
3. How can we tune our approach to improve model quality?
4. How does our technique compose with other model size reduction techniques?

We first compare our technique to the basic hashing trick. The hashing trick performs quite well on our dataset.

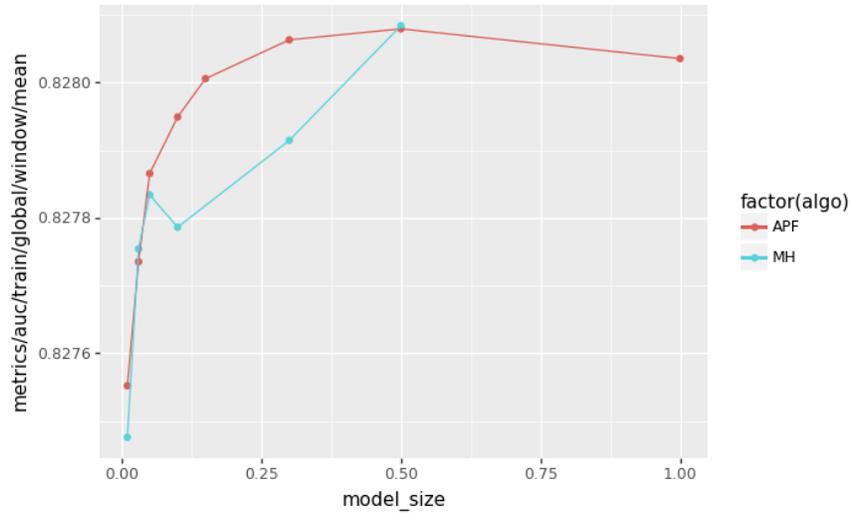

Figure 3: Agglomerative + frequency information in red, hashing trick in blue. 1-4 repetitions for all runs. Higher is better. Agglomerative is most effective vs the hashing trick in the region from 10-30% model size.

## Number of internal tables (cluster dimensions)

Given that our method has the capability to compose with a compression strategy like compositional embeddings, we also experiment with combining our approach with multiple internal embedding tables. Here we find that multiple internal tables does not improve the quality of our approach.

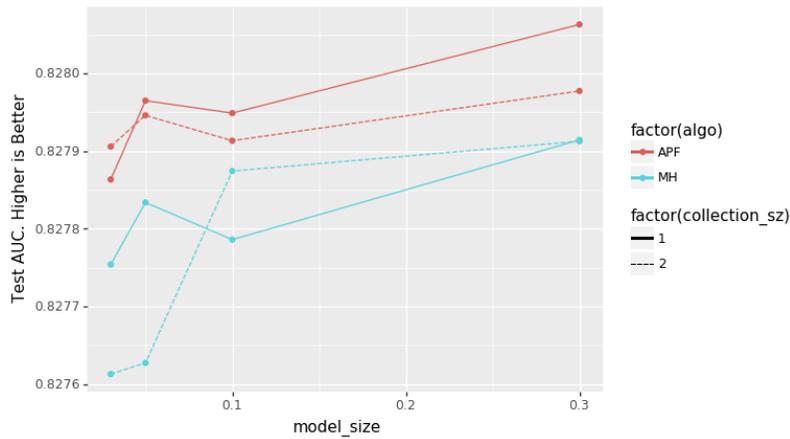

Figure 4: A comparison of a single internal embedding table to two tables. Our approach in red vs the hashing trick in blue. We configure the embedding tables with either one internal table (solid line) or two internal tables (dashed line).

## Comparison to Frequency Based Double Hashing

The frequency based double hashing approach described in [Zheng et al., 2020] is a combination of filtering collisions based on access frequency and multiple internal embedding tables. We implement their solution and experiment on models of 5% size. We find that the frequency based double hashing approach does not improve performance for our model when we retain 1% or 2.5% of vectors in all tables for both algorithms.

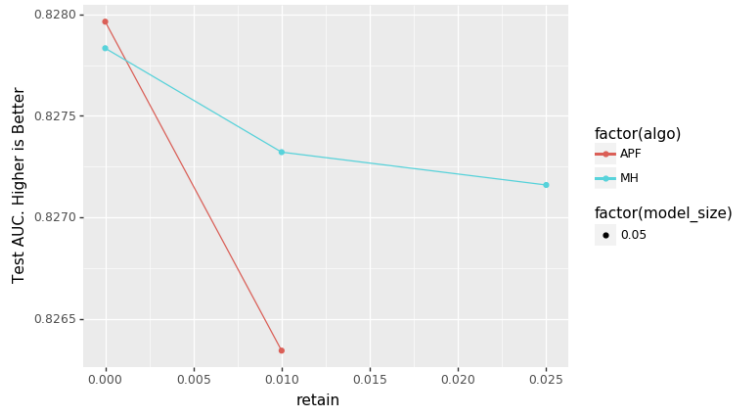

Figure 5: When targeting model sizes of 5%, we found that frequency based double hashing with high frequency retention > 0% performed worse than 0% retention for both our learned hashing and the randomized hashing.

## Access Frequency Information Ablation Study

Given that our algorithm combines both frequency and value information from the teacher model we believe that it is essential to demonstrate that our use of the frequency information is relevant. To further that goal we perform the following ablation study where we disable certain filters from the algorithm. We experiment with configurations with no frequency information, frequency information to weight cluster merges, frequency information for weighting and filtering, and frequency information only for filtering. We demonstrate that frequency information positively improves model quality.

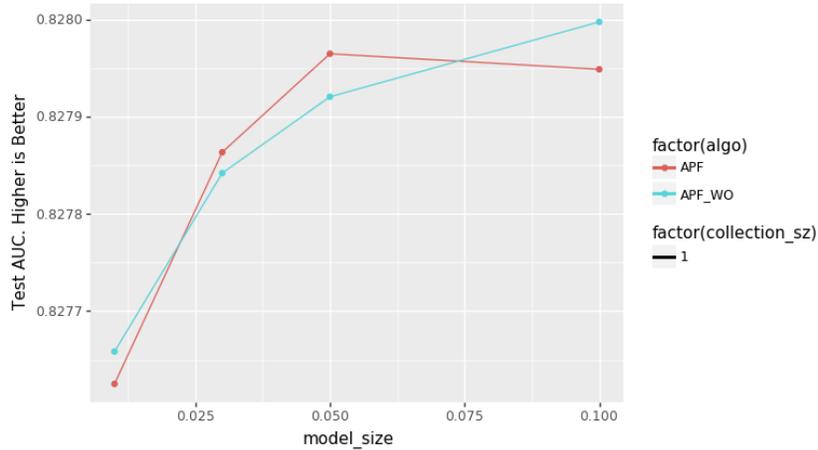

Figure 6: Comparison between weighting and filtering with frequency (red) to only weighting with frequency information (blue). We see that using both weighting and frequency improves performance for the medium model size targets (2%-5%).

We also compare the impact of our algorithm with and without frequency information to when composed with compositional embeddings. We find that while a single internal table outperforms multiple tables in all cases for agglomerative plus frequency information, for the case of two internal table, there is one instance, 15% model size, where agglomerative without frequency information results in a higher quality model.

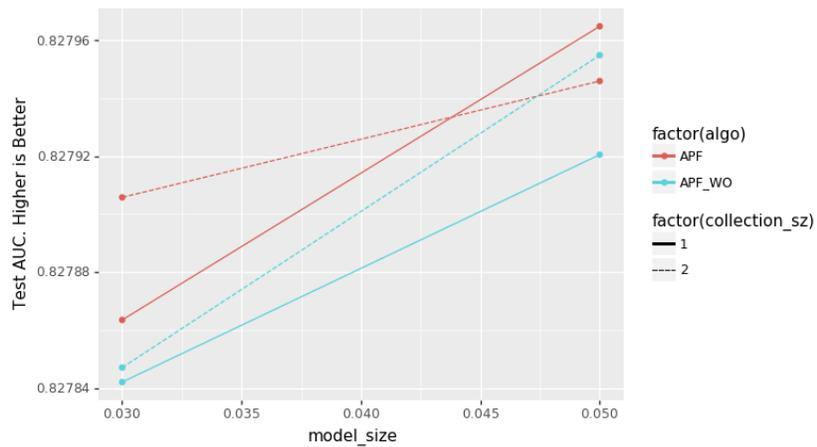

Figure 7: We compare the behavior of our clustering algorithm with and without frequency information used as a filter. Frequency information for weighting and filtering (red), frequency information only for weighting (blue).

# Impact of dataset distribution changes over model days

A fundamental characteristic of our approach is that we use historical data to learn which collisions will have less impact while training the small teacher model. There are three key questions while using historical data:
1. How much historical data should we use?
2. How far should we look back in time?
3. How well do our decisions behave as we move forward in time?

To answer the first question, we train a teacher model that looks backwards over the last week of data and provide it with 750M randomly selected samples and compare it to a teacher model that is trained on 1.5B randomly selected samples from the same time period. We find that the model trained on 1.5B samples performs slightly better than the model trained on 750M. This follows our intuition that more data will allow us to train a better teacher model which will in turn produce a better clustering. Additionally, more data provides more opportunities for low frequency values to be trained.

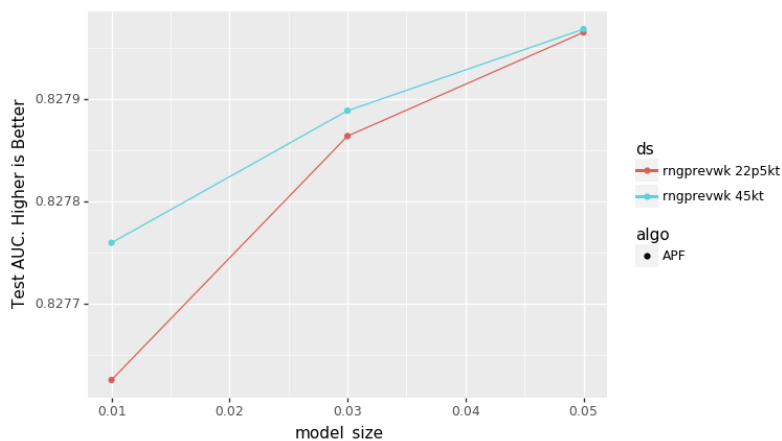

Figure 8: We cluster based on two different teacher models: one with 750M samples (green) and one with 1.5B samples (blue). More samples performs slightly better, especially when the targeted model size is smaller.

Our second experiment addresses question 2. We train one model on 750M samples selected from days one through seven and we train another model on 750M samples from days five through eight. We find that the model trained on 750M samples from the longer time period is slightly better vs the shorter time period. An interesting note here is that the model trained over the shorter time period will contain more samples from the same range as the clustered model training set. This translates to increased performance in the training set, but apparently the model loses some ability to generalize onto later days.

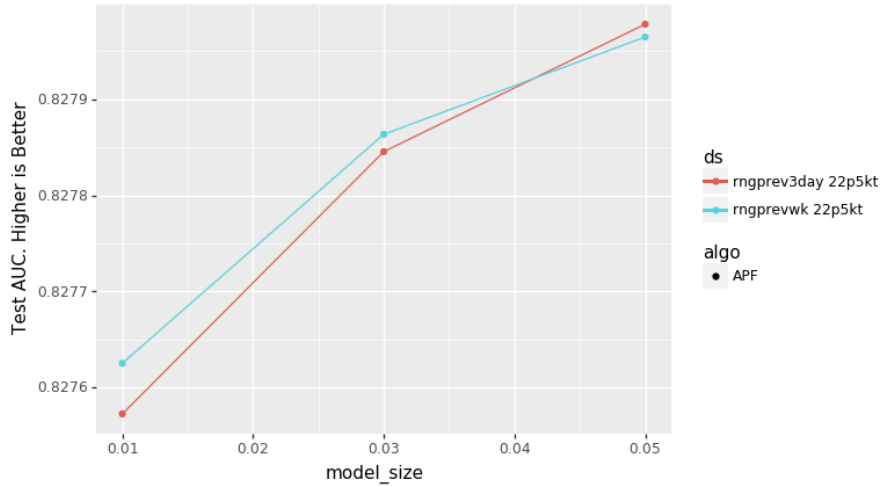

Figure 9: We compare teacher models trained over different days from the training set. The teacher model trained with the same number of samples, but over more days (blue) performs better than the model trained over fewer days (red) for the smaller models.

# Conclusion

We developed and evaluated a new technique for model size reduction on recommendation systems. Our approach learns which categorical ids may be collided with one another and improves model performance vs the hashing trick across all configurations. However, the magnitude of improvement is small and we do not find a scenario in which our approach produces a model with no quality degradation while the hashing trick does. This limits the usefulness of this approach unless small improvements in model quality are critical.